\documentclass[10pt,conference]{IEEEtran}
\IEEEoverridecommandlockouts
\usepackage{cite}
\usepackage[T1]{fontenc}
\usepackage{hyperref}
\usepackage{amsmath,amssymb,amsfonts}
\usepackage{algorithmic}
\usepackage{graphicx}
\usepackage{textcomp}
\usepackage{xcolor}
\usepackage{listings}
\usepackage{microtype}
\usepackage[most]{tcolorbox}
\usepackage{subcaption}
\usepackage[varqu]{inconsolata}
\usepackage{enumitem}
\usepackage{threeparttable}
\usepackage{booktabs}
\usepackage[table]{xcolor}
\usepackage{makecell}
\usepackage{balance}

\newcommand{\rev}[1]{#1}
\newcommand\ans[1]{
\begin{tcolorbox}[size=title, boxrule=0.75pt, colback=green!5, colframe=green!40!black]
\noindent #1
\end{tcolorbox}
}

\definecolor{codebg}{gray}{0.97}
\definecolor{keywordcolor}{rgb}{0.2,0.2,0.7}
\definecolor{commentcolor}{rgb}{0,0.5,0}
\definecolor{stringcolor}{rgb}{0.58,0,0.1}
\definecolor{frameblue}{RGB}{0,112,155}
\lstdefinestyle{customc}{
    backgroundcolor=\color{codebg},
    basicstyle=\ttfamily\scriptsize,
    keywordstyle=\color{keywordcolor}\bfseries,
    commentstyle=\color{commentcolor}\itshape,
    stringstyle=\color{stringcolor},
    numberstyle=\tiny,
    numbers=left,
    stepnumber=1,
    numbersep=8pt,
    xleftmargin=1em,
    showstringspaces=false,
    tabsize=2,
    breaklines=true,
    frame=single,
    captionpos=b,
    language=C,
    morekeywords={uint8_t,static}
}

\def\BibTeX{{\rm B\kern-.05em{\sc i\kern-.025em b}\kern-.08em
    T\kern-.1667em\lower.7ex\hbox{E}\kern-.125emX}}
\begin{document}

\title{What You Trust Is Insecure: Demystifying How Developers (Mis)Use Trusted Execution Environments in Practice
}

\author{
\IEEEauthorblockN{Yuqing Niu,
Jieke Shi*,
Ruidong Han,
Ye Liu,
Chengyan Ma,
Yunbo Lyu,
and David Lo}
\IEEEauthorblockA{
\textit{School of Computing and Information Systems, Singapore Management University, Singapore} \\
\{yuqingniu, jiekeshi, rdhan, yeliu, chengyanma, yunbolyu, davidlo\}@smu.edu.sg}
\thanks{* Corresponding author.}
}

% \and
% \and
% \and
% \and
% \and

\maketitle
\thispagestyle{plain}

% Abstract
\begin{abstract}
Trusted Execution Environments (TEEs), such as Intel SGX and ARM TrustZone, provide isolated regions of CPU and memory for secure computation and are increasingly used to protect sensitive data and code across diverse application domains. However, little is known about how developers actually use TEEs in practice. This paper presents the first large-scale empirical study of real-world TEE applications. We collected and analyzed 241 open-source projects from GitHub that utilize the two most widely-adopted TEEs, Intel SGX and ARM TrustZone. By combining manual inspection with customized static analysis scripts, we examined their adoption contexts, usage patterns, and development practices across three phases. First, we categorized the projects into 8 application domains and identified trends in TEE adoption over time. We found that the dominant use case is IoT device security (30\%), which contrasts sharply with prior academic focus on blockchain and cryptographic systems (7\%), while AI model protection (12\%) is rapidly emerging as a growing domain. Second, we analyzed how TEEs are integrated into software and observed that 32.4\% of the projects reimplement cryptographic functionalities instead of using official SDK APIs, suggesting that current SDKs may have limited usability and portability to meet developers' practical needs. Third, we examined security practices through manual inspection and found that 25.3\% (61 of 241) of the projects exhibit insecure coding behaviors when using TEEs, such as hardcoded secrets and missing input validation, which undermine their intended security guarantees. Our findings have important implications for improving the usability of TEE SDKs and supporting developers in trusted software development.

\end{abstract}


% Introduction
\section{Introduction}\label{sec:intro}

\begin{figure}[t]
\centering
\includegraphics[width=0.95\linewidth]{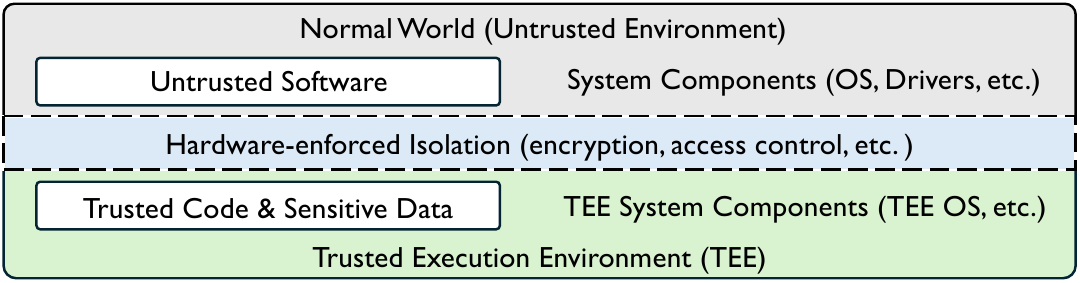}
\caption{Typical architecture of Trusted Execution Environments.}
\label{fig:tee-architecture}
\end{figure}

To safeguard critical computations from potentially compromised software or operating systems~\cite{zheng2021survey}, major hardware vendors such as ARM~\cite{mcgillion2015open} and Intel~\cite{costan2016intel} have introduced {\it Trusted Execution Environments} (TEEs) into their processor architectures~\cite{li2023survey}. A TEE is an isolated region of the CPU and memory that protects sensitive code and data from untrusted software components through hardware-enforced isolation mechanisms such as encryption and access control~\cite{arfaoui2014trusted,zheng2021survey,jauernig2020trusted}. As illustrated in Figure~\ref{fig:tee-architecture}, these mechanisms typically partition the system into two domains: the normal world, which hosts untrusted software and system components, and the TEE, which executes trusted code and manages sensitive data. This isolation ensures that critical computations executed within TEEs remain confidential and tamper-resistant, thereby meeting the increasing demand for trustworthy computing across various domains such as mobile payment~\cite{wang2020building,zheng2016trustpay}, cryptocurrency wallets~\cite{gentilal2017trustzone}, and remote attestation~\cite{10628729}. According to GlobalPlatform's 2024 Annual Report~\cite{globalplatformAnnualReport}, billions of devices worldwide, from smartphones to cloud servers, now incorporate TEEs, and the global TEE market is projected to reach \$22.3 billion by 2033~\cite{growthmarketreportsTrustedExecution}.

% \begin{figure}[t!]
% \begin{minted}[fontsize=\scriptsize,bgcolor=gray!5,breaklines]{diff}
% // GitHub Project: shuaifengyun/basicAlg_use (commit 327f23d)
% // File: core/tee/crypto_ta_pbkdf2.c
% @@ -261,6 +261,9 @@ void g_CryptoTaPbkdf_PBKDF2(..., int dkLen)
% // Improper TEE usage: missing input validatio
% -    TEE_MemMove(output, resultBuf, dkLen);
% // Fix: Added input validation to prevent buffer overflow
% +    if (dkLen > 512)
% +        return TEE_ERROR_BAD_PARAMETERS;
% +    TEE_MemMove(output, resultBuf, dkLen);
%      return TEE_SUCCESS;
% \end{minted}
% \captionof{listing}{Example of improper TEE usage in a cryptographic library on GitHub (\texttt{shuaifengyun/basicAlg\_use}), where a security weakness arises from missing input validation for data originating from the untrusted normal world. This issue was detected and fixed by a recent study~\cite{ma2025diting}.}
% \label{lst:pbkdf2-vuln-diff}
% \end{figure}

Hardware vendors, who are the primary implementers of TEEs, typically disclose only limited details of their internal designs to software developers. Instead, they provide software development kits (SDKs), such as the Intel Software Guard Extensions (SGX) SDK~\cite{intel_sgx_sdk} and the Open Portable (OP)-TEE framework~\cite{optee_docs}, to standardize and facilitate TEE integration in user applications. However, developing TEE-based software presents unique challenges, particularly in managing encrypted memory and controlling communication between trusted and untrusted environments~\cite{li2019teev,paju2023sok}, which demands considerable security expertise.
While prior studies have shown that developers often exhibit insecure practices in cryptographic and other security-critical contexts, such as misusing Application Programming Interfaces (APIs)~\cite{nadi2016jumping,arzt2015towards,mousavi2025detecting,galappaththi2024empirical}, little is known about how such issues manifest in the uniquely complex setting of TEE development.
Existing studies have primarily focused on analyzing TEE-related mobile binaries~\cite{bove2024large} or systematizing academic literature on TEE use cases~\cite{paju2023sok}, leaving a gap for an empirical investigation that addresses a key question:
\textit{Do developers adhere to the intended usage patterns of official TEE SDKs, or do practical constraints cause them to deviate from recommended practices?}

To bridge this gap, we present the first large-scale quantitative and qualitative investigation of how TEEs are used in real-world software projects. Our study focuses on the two most widely-adopted TEEs, Intel SGX~\cite{costan2016intel} and ARM TrustZone~\cite{mcgillion2015open}, which together account for the majority of publicly available TEE development efforts~\cite{li2023survey}. Using the GitHub Search API with TEE-related keywords, we followed prior work~\cite{humbatova2020taxonomy,borges2016understanding} to iteratively refine our search queries by adjusting repository size filters, thereby maximizing the coverage of relevant projects. This process yielded 1,547 open-source repositories created or updated between January 2011 and May 2025. After excluding tutorial and documentation-only repositories, as well as unpopular ones with fewer than 10 stars, we manually inspected each repository to confirm the presence of actual TEE implementations. The resulting curated dataset consists of 241 repositories selected for detailed analysis (see Section~\ref{sec:methodology}). We then combine manual annotation and static analysis to answer the following research questions (RQs):

\begin{itemize}[leftmargin=*]
    \item \textbf{RQ1.} In which domains are TEEs adopted in practice?
    \item \textbf{RQ2.} What are the prevailing patterns in TEE development?
    \item \textbf{RQ3.} What insecure practices and pitfalls are prevalent in real-world TEE projects?
\end{itemize}

To understand where TEEs are applied in practice, we categorized all 241 repositories into 8 primary application domains (e.g., AI, blockchain, or secure storage) by analyzing TEE-specific API calls using our customized static analysis scripts. The results show that TEEs are increasingly adopted across diverse domains, yet this trend diverges from prior academic hotspots that emphasize security-critical blockchain and cryptographic systems~\cite{gentilal2017trustzone,li2023survey,liu2022extending,zheng2016trustpay,zheng2021survey}. In practice, Internet-of-Things (IoT) device security dominates (30\%), whereas blockchain-related projects account for only 7\%. Interestingly, AI model protection~\cite{mo2021ppfl,mo2020darknetz} has emerged as a fast-growing category (12\%) with many projects appearing in recent years, signaling a shift toward protecting machine learning models and data within TEEs and aligning with the broader trend of secure AI development~\cite{zhang2022teeslice,lo2023trustworthy}.

We also categorized the projects based on the functionality implemented or migrated into TEEs. Cryptographic operations (68.2\%) represent the most prevalent functionality implemented within TEEs, which is expected. However, many projects (35.6\% of Intel SGX and 27.1\% of ARM TrustZone projects) reimplement standard cryptographic operations like encryption algorithms, instead of relying on official SDK APIs. Such reimplementations can introduce logical flaws and side-channel vulnerabilities~\cite{ma2025diting,yuan2024hypertheft,joy2024physicalsoftwarebasedfault}, while also revealing the limited usability of existing SDKs, where unclear documentation and opaque technical details often lead developers to bypass safe and off-the-shelf APIs. Furthermore, our manual inspection of the source code revealed that 25.3\% (61 of 241) of the analyzed projects exhibit insecure coding practices, such as missing input validation or hardcoded secrets. These findings suggest that maintaining sound security principles within the uniquely complex TEE environments remains challenging, and even security-conscious developers may turn TEEs into attack surfaces rather than trusted boundaries. We therefore discuss implications for TEE toolchain designers and the broader secure-systems community to better align TEE capabilities and SDK design with real-world developer practices and to foster safer and more accessible TEE development.

To summarize, our key contributions are as follows:

\begin{itemize}[leftmargin=*,topsep=0pt]
    \item \textbf{Dataset.} We conduct the first large-scale empirical study of TEE usage in real-world projects, contributing a curated dataset of 241 open-source repositories with detailed annotations of TEE-related implementations and functionalities.
    \item \textbf{Findings.} Our analysis reveals that real-world TEE usage diverges from long-held research hotspots and its secure design intents: TEEs are predominantly used for IoT security (30\%) rather than for blockchain security (7\%). Moreover, 32.4\% of the projects bypass the official SDK cryptographic APIs, and 25.3\% exhibit insecure coding practices.
    \item \textbf{Implications.} We identify key usability gaps in current TEE toolchains and propose actionable implications for SDK designers and the broader secure systems community.
\end{itemize}

% The rest of this paper is organized as follows. Section~\ref{sec:background} introduces the basics of TEEs and related work. Section~\ref{sec:methodology} presents our empirical methodology, including project annotation and analysis workflow. Section~\ref{sec:findings} reports our findings and we discuss lessons learned and future directions in Section~\ref{sec:discussion}. Section~\ref{sec:threats} outlines potential threats to validity, and Section~\ref{sec:conclusion} concludes the paper.

\section{Background \& Related Work}\label{sec:background}

\subsection{Trusted Execution Environments}
\label{sec:tee-background}
\begin{lstlisting}[
    style=customc,
    float=t!,
    caption={Simplified cross-world invocation in a TEE. Only one line of code in the normal world invokes a secure hash function inside the TEE.},
    label=lst:tee-example,
    morekeywords={Input, Result, TEE_Result, TEE_Success}
]
// normal_app.c  (Normal world)
#include <tee_client_api.h>   // TEE SDK header
Input data = "hello world!";
Result output;
tee_invoke("secure_hash", data, &output);   // Cross into TEE

// secure_app.c  (TEE)
#include <tee_internal_api.h>
TEE_Result secure_hash(Input data, Result *output) {
    TEE_DigestDoFinal(hash_ctx, data.buf, data.size, output->buf, &output->size);
    return TEE_Success;
}
\end{lstlisting}

Trusted Execution Environments protect sensitive code and data from untrusted operating systems and privileged software through hardware-level isolation mechanisms such as memory encryption and access control~\cite{zheng2021survey,li2023survey}. The two most widely-deployed TEEs are Intel SGX~\cite{costan2016intel} and ARM TrustZone~\cite{mcgillion2015open}, designed for the Intel x86 and ARM architectures, respectively. Although they differ in implementation details, both share a common design principle of partitioning the system into two domains: the normal world, which runs untrusted applications, and the TEE, which executes trusted code (called an {\it enclave} in Intel SGX and a {\it secure world} in TrustZone). In TrustZone, these trusted components are also referred to as Trusted Applications (TAs). Both platforms provide SDKs that enable developers to bridge the two domains with minimal effort. Listing~\ref{lst:tee-example} illustrates a simplified interaction, where an untrusted application in the normal world invokes a secure hash function implemented inside the TEE via a vendor-provided API. In this example, developers write only a few lines of code to call the secure function, while the underlying SDK automatically manages cross-world communication.

Note that the two TEEs studied in this work adopt distinct API designs within their SDKs. In the Intel SGX SDK, developers define the boundary between trusted and untrusted code using Enclave Definition Language (EDL) files, which are compiled into bridge stubs that manage all communication between the enclave and the application. In contrast, OP-TEE, the reference SDK for ARM TrustZone, follows a different workflow in which developers implement both the TA running inside the TEE and the client program in the normal world. The two components communicate through SDK-provided APIs, typically prefixed with \texttt{TEE\_} or \texttt{TEEC\_}. Although API names, conventions, and even the terminology used to refer to TEEs vary across platforms, we include all such variants in our search process. We also leverage these platform-specific API call patterns as reliable indicators when identifying TEE-related implementations and functionalities.

Recent years have witnessed a growing number of TEE applications, spanning secure payment systems~\cite{wang2020building,zheng2016trustpay}, blockchain protocols~\cite{li2019teev,liu2022extending}, and privacy-preserving AI inference~\cite{mo2020darknetz,mo2021ppfl,yuan2024hypertheft}. However, the security guarantees of TEEs ultimately depend on how developers design, implement, and interact with enclave or TA code. In practice, the SDKs' abstraction of low-level operations and the limited disclosure of implementation details may cause developers to misuse TEEs or even introduce new vulnerabilities~\cite{ma2025diting,cerdeira2020sok,joy2024physicalsoftwarebasedfault}, motivating our empirical study of real-world TEE usage patterns and pitfalls.

\subsection{Research Landscape on TEE Usability and Adoption}

The main body of studies on TEEs can be broadly divided into two categories: system abstractions that simplify TEE development and domain-specific applications that leverage TEE capabilities. The first line of work focuses on building frameworks that abstract low-level TEE operations~\cite{shinde2017panoply,tsai2017graphene,paju2023sok}. Representative examples include Graphene-SGX~\cite{tsai2017graphene}, which enables Linux-compatible system calls inside enclaves; SCONE~\cite{arnautov2016scone}, which supports containerized enclave execution with automatic encryption and attestation; and Keystone~\cite{lee2020keystone}, which extends TEE support to RISC-V platforms with customizable enclaves. While these systems lower the entry barrier for developers, they also introduce additional complexity and performance overhead, and none have achieved widespread adoption in practice compared to the official SDKs.
The second line of work explores domain-specific use cases that leverage TEE capabilities, such as secure data analytics~\cite{yang2023vedb}, key management~\cite{gentilal2017trustzone,han2025autotee}, secure networking~\cite{li2019teev,liu2022extending}, privacy-preserving machine learning~\cite{mo2020darknetz,mo2021ppfl,yuan2024hypertheft}, and more~\cite{kim2019shieldstore,yang2023vedb}. These studies demonstrate the feasibility of secure computation within TEEs across various domains but often rely on handcrafted or highly specialized implementations that may not reflect how developers use TEEs in everyday practice.

Complementing these system-building efforts, several surveys and systematization papers have aimed to capture the broader TEE research landscape. Paju et al.~\cite{paju2023sok} provide a comprehensive literature review that categorizes use cases, SDKs, and usability challenges. Although informative, their work focuses on the academic state of TEE research rather than real-world development practices. Bove et al.~\cite{bove2024large} analyze TEE usage in Android applications and find that direct API usage is rare; however, their analysis targets mobile binaries rather than open-source projects with actual implementations, still leaving a gap in understanding developer practices.

Prior studies in the software engineering and security communities have shown that even experienced developers frequently misuse security-critical APIs~\cite{fahl2012eve,nadi2016jumping,meng2018secure,wijayarathna2019using,gorski2016towards}. However, development within TEEs, which may be particularly error-prone due to unique challenges such as cross-world communication and encrypted memory management, remains largely understudied. This gap motivates our empirical study to provide a more comprehensive understanding of real-world TEE development practices.

% Methodology
\section{Data Collection \& Labeling}\label{sec:methodology}
Here we describe how we collect, filter, and label TEE-related projects to form the basis of our empirical analysis. The scope of this study is defined along two dimensions: the targeted TEE platforms and the sources of analyzed projects. As discussed in Section~\ref{sec:tee-background}, we focus on the two most widely-adopted TEEs, Intel SGX~\cite{costan2016intel} and ARM TrustZone~\cite{mcgillion2015open}, which together dominate real-world deployments. Other TEEs, such as RISC-V Keystone~\cite{lee2020keystone}, are excluded due to their limited adoption and the lack of publicly available implementations. We specifically target open-source projects hosted on GitHub, the largest collaborative development platform and a rich source of real-world software artifacts~\cite{cosentino2016findings}. No restrictions are imposed on programming languages or project timelines, allowing us to capture a comprehensive snapshot of TEE usage across diverse domains and development practices.

With this scope in mind, our data collection process comprises three key stages, as detailed below.

\vspace{0.1cm}
\noindent
\textbf{1) Keyword-Based Repository Search.}
We constructed a comprehensive Boolean search query to retrieve candidate repositories from GitHub. The search query incorporated two conceptual components: \emph{TEE terminology} and \emph{development artifacts}. Specifically, we used the following Boolean expression:
\begin{tcolorbox}[colback=white, colframe=black, boxrule=0.5pt, arc=1mm, left=1mm, right=1mm, top=1mm, bottom=1mm]
{\small
\begin{verbatim}
("Intel SGX" OR "SGX enclave" OR "enclave" OR "TrustZone"
OR "ARM TrustZone" OR "OP-TEE" OR "Trusted Application"
OR "secure world" OR "TEE")
                        AND
("ecall" OR "ocall" OR ".edl" OR "sgx_create_enclave"
OR "TEEC_InvokeCommand" OR "TA_OpenSessionEntryPoint"
OR "trusted app")
\end{verbatim}
}
\end{tcolorbox}
Note that the first component captures diverse terms referring to TEEs and their related concepts, whereas the second component targets keywords associated with TEE-specific APIs and development artifacts, such as EDL files and \texttt{ECALL}/\texttt{OCALL} functions for cross-world communication in Intel SGX, as well as OP-TEE API calls like \texttt{TEEC\_InvokeCommand} for invoking code in the secure world. Together, these two components ensure that the search results predominantly reflect actual TEE development activities rather than incidental mentions of TEEs.

The query was executed using the GitHub Search API. Since the GitHub Search API returns at most 1,000 results per query, we adopted an iterative search strategy following the approach of Humbatova et al.~\cite{humbatova2020taxonomy}. Specifically, we varied the \texttt{size:min..max} parameter across multiple queries to circumvent this limitation and capture a broader range of repositories. Each query covered a distinct file size range from 0 to 384,000 bytes, which is the maximum allowed by GitHub, with a step size of 250 bytes to ensure comprehensive coverage of relevant projects. Using this strategy, we retrieved 3,452 files related to OP-TEE, mapped to 875 unique repositories, and 2,438 files for Intel SGX, mapped to 672 unique repositories. Each file was associated with its corresponding repository through metadata, and results from all query sources were merged to form a unified dataset. The resulting repositories were created or last updated between January 2011 and May 2025, covering more than 14 years of TEE development and yielding a total of 1,547 unique repositories.

\vspace{0.1cm}
\noindent
\textbf{2) Filtering Based on Popularity and Relevance.}
To ensure both the practical significance and technical quality of the collected repositories, we applied two filtering criteria:
\begin{itemize}[leftmargin=*]
    \item \textbf{Popularity Filter:} We retained only repositories with at least 10 GitHub stars. This threshold helps remove inactive or trivial projects while preserving those that have received community attention or sustained maintenance. Prior studies have shown that popularity metrics such as stars are effective indicators of project maturity and relevance~\cite{cosentino2016findings,lyu2025my}. \rev{This step removed a substantial number of 1,250 repositories.}
    \item \textbf{Technical Relevance Filter:}
    \rev{We manually screened the remaining repositories to exclude those explicitly marked as tutorials, homework, or simple ``hello-world'' demos in their README files or descriptions. Although such projects often contain executable code, they lack actual application logic. Therefore, this step focused on filtering out toy examples to ensure that only technically meaningful projects were included in our dataset, excluding 40 repositories.}
\end{itemize}
\noindent
\vspace{0.1cm}
\noindent
\textbf{3) Automated Screening for TEE Implementations and Manual Verification.}
Our final filtering step ensured that each collected repository contained genuine TEE-specific implementations rather than superficial mentions of TEE-related terms. To achieve this, we combined customized static analysis scripts with manual verification to confirm the presence of actual TEE code implementing concrete functionalities. We developed a set of static analysis scripts that applied keyword matching, prefix detection, and directory pattern recognition to identify TEE-specific artifacts (e.g., EDL files and TA source files) by scanning each repository for several indicators, including: (1) dedicated source directories such as \texttt{ta/}, \texttt{enclave/}, or \texttt{sgx/}, which typically contain trusted code components or enclave build artifacts; (2) platform-specific build or configuration files such as \texttt{Enclave.config.xml} or \texttt{use\_ta.mk}, which indicate that the project is intended to compile and execute TEE components rather than include placeholder references; and (3) the presence of TEE-specific API calls (e.g., \texttt{TEEC\_}, \texttt{TEE\_}, or \texttt{sgx\_} prefixes), which reflect direct interaction with enclaves or trusted applications and serve as strong evidence of genuine TEE usage.

Following the automated screening, we manually reviewed each repository to confirm the presence of authentic TEE implementations. \rev{In contrast to the lightweight check in Step 2, which considered only non-code information such as GitHub stars and project descriptions, this verification involved an in-depth inspection of the source code to determine whether TEE APIs were functionally integrated into the application's core logic.} This process involved some subjective judgment, particularly for repositories with complex code that required domain expertise to assess genuine TEE usage. To mitigate potential bias, each repository was independently reviewed by two authors to ensure consistency in labeling. The two authors achieved an ``almost perfect'' level of agreement~\cite{viera2005understanding}, with a Cohen's $\kappa$ of 0.92. \rev{Disagreements arose in only 8 cases and were resolved through discussion and consensus.}

\rev{With the above automated screening and manual validation, we further excluded 16 repositories.} Finally, we curated a dataset of 241 projects, comprising 92 using OP-TEE and 149 using Intel SGX. This dataset serves as the foundation for our subsequent empirical analysis.

\section{Empirical Results and Findings}\label{sec:findings}

In this section, we present our findings based on an analysis of 241 real-world TEE projects. The results are organized around the three research questions introduced in Section~\ref{sec:intro}, combining quantitative statistics with qualitative insights.

\subsection{RQ1: In which domains are TEEs adopted in practice?}\label{sec:findings-domain}

Although prior studies have identified TEEs as key enablers of secure computing in sectors such as mobile payment and blockchain~\cite{gentilal2017trustzone,liu2022extending}, their adoption in real-world applications remains insufficiently understood. Industrial usage may have shifted toward emerging areas, diverging from the traditional security-critical domains emphasized in academic research. By systematically mapping the domains of TEE usage, we aim to uncover where TEEs are truly applied in practice and how these trends have evolved, thereby guiding future academic studies toward greater practical relevance.

\vspace{0.1cm}
\noindent
\textbf{Categorization Process.}
Using the curated dataset of 241 TEE projects, we manually categorized each repository according to its primary application domain. We first examined project documentation (e.g., \texttt{README.md}) and repository descriptions. When this information was insufficient to determine the domain of TEE usage, we conducted source code–level inspections supported by customized static analysis scripts that detect TEE-specific APIs and configurations, as described in Section~\ref{sec:methodology}. The static analysis identified TEE code folders and API prefixes (e.g., \texttt{TEEC\_} or \texttt{sgx\_}) to locate functionalities implemented inside the TEE, which were then manually verified.

To derive a comprehensive taxonomy of TEE application domains, we adopted the \textit{open card sorting} method~\cite{begel2014analyze,cruzes2011recommended,stol2016grounded,yang2023users}, a qualitative technique widely used for categorizing software artifacts when predefined categories are unavailable. This approach allows categories to emerge through iterative grouping and refinement, which suits our context where TEE projects' application domains are diverse and domain boundaries are often ambiguous. In our card sorting, each project was assigned to the domain that best represents its core functionality. For example, a repository implementing a secure communication protocol for IoT devices was classified under \textit{IoT Device Security} but not under \textit{Network Security}, even if it also involves network encryption components. Using the open card sorting method and our manual inspection of TEE usage domains, we performed categorization in three steps:

\begin{itemize}[leftmargin=*]
\item \textbf{Initial Independent Categorization.} Two authors reviewed all repositories independently, examining documentation and source code to identify their primary domains, \rev{aiming to enhance the reliability of the classification.}
\item \textbf{Group Discussion and Consensus Building.} The two authors then discussed their initial categorizations, resolved discrepancies, and refined the emerging taxonomy through consensus. New categories were also created when necessary.
\item \textbf{Final Review and Validation.} A third author then joined, and the three authors conducted a comprehensive review of all categorizations, revisiting ambiguous cases and cross-checking each project's classification against its stated goals and implemented functionalities.
\end{itemize}

The entire categorization process took approximately 60 hours in total.

\begin{table}[t!]
\centering
\caption{Distribution of TEE Projects by Application Domain}
\label{tab:project-domains}
\rowcolors{2}{gray!10}{white} % 从第二行开始，奇数行背景为浅灰色 (10% 灰)，偶数行背景为白色
\begin{tabular}{lcc}
\toprule[1.5pt]
\textbf{Category}  & \textbf{Count} & \textbf{Pct.\%} \\
\midrule[0.8pt]
IoT Device Security & 71 & 30\% \\
Privacy-Preserving Computing & 46 & 19\% \\
Network Security \& Encrypted Communication & 42 & 17\% \\
AI Model Protection & 28 & 12\% \\
Database Security \& Secure SQL & 20 & 8\% \\
Blockchain \& Cryptocurrency Security & 17 & 7\% \\
U-Boot \& Secure Boot & 10 & 4\% \\
Others & 7 & 3\% \\
\bottomrule[1.5pt]
\end{tabular}
\end{table}

\vspace{0.1cm}
\noindent
\textbf{Categorization Results.}
We identified 7 major domains of TEE applications, and a small number of projects (about 3\%) that did not clearly fit into these categories were grouped under \textit{Others}, as shown in Table~\ref{tab:project-domains}, along with their project counts and percentage shares. Below, we briefly describe each domain:
\begin{itemize}[leftmargin=*]
\item \textbf{IoT Device Security (30\%)}: Projects that secure IoT devices through encrypted storage, device authentication, and firmware protection. Common applications include routers, set-top boxes, and single-board computers (SBCs).
\item \textbf{AI Model Protection (12\%)}: Projects that deploy AI or deep learning models inside TEEs to safeguard model confidentiality and data privacy during inference or training.
\item \textbf{Blockchain and Cryptocurrency Security (7\%)}: Projects leveraging TEEs for blockchain transaction integrity and privacy, such as confidential smart contracts~\cite{liu2025towards}.
\item \textbf{Network Security and Encrypted Communication (17\%)}: Projects securing network data transmission using TEEs, including Transport Layer Security (TLS) endpoints~\cite{apostolopoulos1999transport} and encrypted communication channels.
\item \textbf{Database Security \& Secure SQL (8\%)}: Projects running databases (e.g., SQLite) within TEEs to protect sensitive data and ensure secure query execution.
\item \textbf{Privacy-Preserving Computing (19\%)}: Projects that use TEEs for encryption-based computation or analytics, often integrating cryptographic primitives for data confidentiality.
\item \textbf{U-Boot \& Secure Boot (4\%)}: Projects that integrate TEEs into the bootloader to enforce trusted boot sequences and ensure system integrity from startup.
\item \textbf{Others (3\%)}: Projects that do not clearly fall into the above categories, including experimental frameworks, or multi-domain systems that span several use cases.
\end{itemize}

\begin{figure}[t!]
\centering
\includegraphics[width=0.5\textwidth]{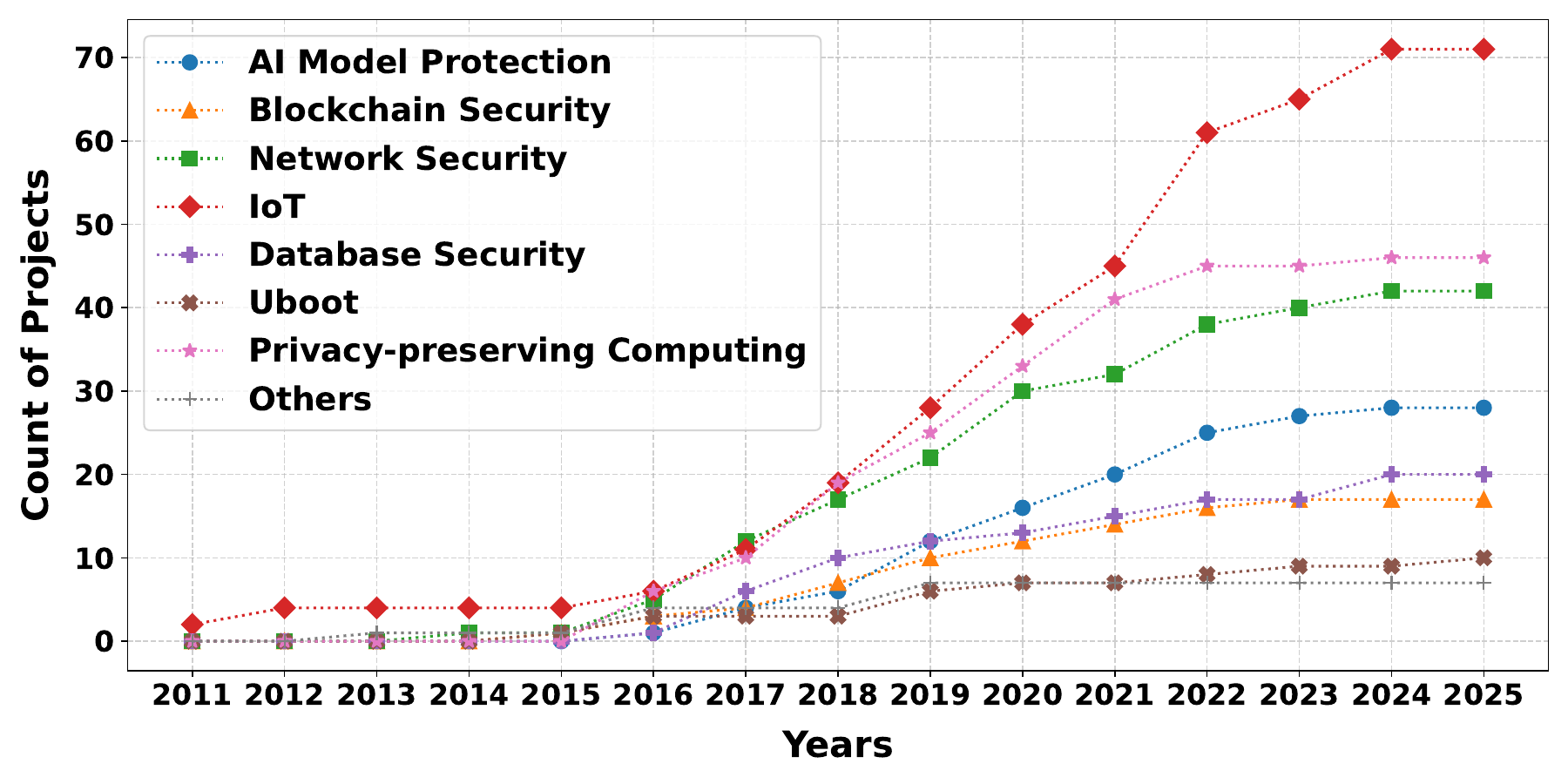}
\caption{Cumulative growth of TEE projects by domain (2011-2025).}
\label{fig:domain-timeline}
\end{figure}

\vspace{0.1cm}
\noindent
\textbf{Key Trends and Insights.}
The domain distribution of TEE projects reveals several notable patterns. First, although prior research has largely emphasized blockchain and cryptographic systems as primary TEE applications~\cite{gentilal2017trustzone,zheng2021survey,li2023survey,zheng2016trustpay}, these domains together account for only 7\% of real-world projects. In contrast, \textit{IoT Device Security} dominates (30\%), where TEEs are primarily used to ensure device integrity and enable secure updates. Second, the prominence of \textit{Privacy-Preserving Computing} (19\%) and \textit{Network Security \& Encrypted Communication} (17\%) indicates that developers increasingly employ TEEs not only for ``isolation'' but also to build secure computation and communication services within TEEs. This shift reflects the growing maturity of TEEs in supporting complex software stacks and the increasing demand for data protection driven by real-world regulatory pressures (e.g., GDPR~\cite{zhang2023pacta,istvan2020towards}).

However, an interesting observation is the growing adoption of \textit{AI Model Protection} (12\%), a relatively new domain that signifies a meaningful convergence between academic research on confidential AI~\cite{mo2020darknetz,mo2021ppfl,yuan2024hypertheft} and practical developer needs. Prior research works like DarkneTZ~\cite{mo2020darknetz} and PPFL~\cite{mo2021ppfl} \rev{were among the first to demonstrate the feasibility of} embedding AI inference within TEEs to prevent model theft and data leakage. \rev{The emergence of this domain in our dataset (12\%) suggests that privacy risks initially highlighted by academia are now translating into practical developer needs for securing machine learning workloads.} We also observe that \textit{U-Boot \& Secure Boot} projects (4\%) are less common, given TEEs' original design goal of ensuring platform integrity. This trend may reflect the increasing complexity of modern systems, where secure boot is often only one component of a broader security architecture, and developers may prioritize higher-level applications over low-level boot integrity.

We also analyzed the temporal evolution of TEE usage across these domains. Figure~\ref{fig:domain-timeline} shows the cumulative growth of projects in each domain from 2011 to 2025. Early projects (2011–2016) primarily focused on \textit{IoT Device Security} and \textit{Secure Boot}, reflecting TEE's initial role in firmware protection. After that, adoption broadened to \textit{Privacy-Preserving Computing} and \textit{Network Security}, enabled by the maturation of TEE SDKs and growing interest in encrypted communication and attestation. After 2019, \textit{AI Model Protection} rapidly emerged, coinciding with the rise of confidential AI~\cite{mo2021ppfl}, and has recently become one of the most popular application domains. Overall, this trend illustrates a shift from low-level device protection toward data- and AI model–centric confidentiality, marking the evolution of TEEs from hardware isolation mechanisms to a general-purpose foundation for secure computation.

\tcbset{
  rqsummary/.style={
    colback=gray!5!white,
    colframe=black,
    fonttitle=\bfseries,
    boxrule=0.8pt,
    arc=2pt,
    left=4pt,
    right=4pt,
    top=4pt,
    bottom=4pt
  }
}

% ：
% \begin{tcolorbox}[colback=gray!10!white, colframe=frameblue,
% title=RQ1 Summary - Application Domains of TEE Usage,
% boxsep=4pt,          % 减少文本与边框的距离（默认约为 4pt）
%   left=2pt, right=2pt, % 控制左右内边距
%   top=2pt, bottom=2pt, % 控制上下内边距
%   fontupper=\small     % 调整文字大小
%   ]
% TEE adoption is dominated by \textit{IoT Device Security} (30\%), \textit{Privacy-Preserving Computing} (19\%), and \textit{Network Security} (17\%), whereas blockchain accounts for only 7\%. Emerging domains like \textit{AI Model Protection} (12\%) signal a shift from hardware isolation toward data- and AI-centric secure computation.
% \end{tcolorbox}

\ans{\textbf{RQ1 Summary:} Real-world TEE adoption is dominated by \textit{IoT Device Security} (30\%), while traditional research hotspots like blockchain security account for only 7\%. Emerging domains like \textit{AI Model Protection} (12\%) indicate a clear shift in TEE usage from low-level device protection toward data- and AI-centric secure computation.}

% TEE adoption spans seven domains, with \textit{IoT Device Security} (30\%), \textit{Privacy-Preserving Computing} (20\%), and \textit{Network Security} (19\%) being the most prevalent. Traditional research hotspots such as blockchain and cryptographic systems account for only 7\% of projects, while emerging domains like \textit{AI Model Protection} (12\%) are rapidly growing. Overall, the trend highlights a shift from hardware-level isolation toward data- and AI model-centric secure computation.

\subsection{RQ2: What are the prevailing patterns in TEE development?}\label{sec:findings-devpractices}

While RQ1 identifies the domains where TEEs are adopted, understanding \textit{how} developers integrate TEEs within software systems is equally crucial for revealing their practical role in real-world development. It remains unclear which functionalities developers migrate into TEEs and how these migrations are performed. By systematically categorizing the functionalities implemented inside TEEs and conducting a fine-grained manual analysis of their implementation details, we aim to uncover common usage patterns and identify potential misalignments with recommended practices.

\begin{table}[t!]
\centering
\caption{Distribution of Functionalities Implemented Inside TEEs}
\label{tab:functionalities}
\rowcolors{2}{gray!10}{white} % 从第二行开始，奇数行背景为浅灰色 (10% 灰)，偶数行背景为白色
\begin{tabular}{lc}
\toprule[1.5pt]
\textbf{Functionality Category} & \textbf{Projects (\%)} \\
\midrule[0.8pt]
Cryptographic Primitives & 68.2\% \\
Secure Storage  & 43.6\% \\
Remote Attestation Support & 28.7\% \\
Secure Communication / I/O Handling & 16.7\% \\
AI Model Inference & 14.0\% \\
Access Control  & 11.2\% \\
\bottomrule[1.5pt]
\end{tabular}
\end{table}

\vspace{0.1cm}
\noindent
\textbf{Categorization Process.}
Our labeling process for all 241 repositories aimed to identify implemented functionalities executed inside TEEs. It followed the manual and static analysis–assisted screening method introduced in Section~\ref{sec:methodology}. Specifically, we used our customized static analysis scripts that applied keyword matching, prefix detection, and directory pattern recognition to identify (i) the presence of TEE-specific API calls (e.g., \texttt{TEE\_}, \texttt{sgx\_}, \texttt{ocall\_}) and (ii) the location of source files within TEE-specific directories (e.g., \texttt{ta/}, \texttt{enclave/}, \texttt{sgx/}). Using the detailed code context and the scanned locations of these TEE-specific artifacts, we then manually identified and tagged all functions implemented within TEEs, classifying them by functional purpose. The tagging process adopted the open coding method described in Section~\ref{sec:findings-domain}, involving iterative and collaborative refinement to construct a consistent vocabulary of functional categories. Each Function was independently reviewed by two authors to ensure consistency and accuracy, with disagreements resolved through discussion. A third author subsequently conducted a final validation review to reach consensus. This categorization took approximately 80 hours in total.

\vspace{0.1cm}
\noindent
\textbf{Categorization Results.}
Table~\ref{tab:functionalities} summarizes the distribution of functionalities across all 241 projects, showing the proportion of repositories that implement \textit{at least one} function from each category. This table reflects how frequently each functionality is adopted in practice. Our analysis shows that the functionalities implemented inside TEEs can be classified into 6 groups:
\begin{itemize}[leftmargin=*]
\item \textbf{Cryptographic Primitives (68.2\%)}: Core operations such as symmetric/asymmetric encryption, digital signatures, and key management, often reimplemented manually using OpenSSL~\cite{opensslOpenSSL} instead of vendor-provided SDKs.
\item \textbf{Secure Storage (43.6\%)}: Confidential configuration storage, credential data sealing, and TEE-backed file systems designed to prevent unauthorized access.
\item \textbf{Remote Attestation (28.7\%)}: Generation and verification of attestation quotes to establish trust with external parties.
\item \textbf{AI Model Inference (14.0\%)}: Executing selected layers or full neural network models inside TEEs to safeguard proprietary parameters or training data.
\item \textbf{Secure Communication and I/O Handling (16.7\%)}: Implementing secure channel endpoints like TLS~\cite{apostolopoulos1999transport}, message integrity verification, or input/output filtering.
\item \textbf{Access Control and Decision Making (11.2\%)}: Policy enforcement, user authentication, and secure session management within TEE applications.
\end{itemize}

\begin{table}[t!]
\centering
\begin{threeparttable}
\caption{Cross-Domain Distribution of High-Level TEE Functionalities}
\label{tab:function-cross-domain}
\rowcolors{2}{gray!10}{white} % 从第二行开始，奇数行背景为浅灰色 (10% 灰)，偶数行背景为白色
\begin{tabular}{lccccccc}
\toprule[1.5pt]
\textbf{Category} & \textbf{AI} & \textbf{IoT} & \textbf{BC} & \textbf{DB} & \textbf{PP} & \textbf{NC} & \textbf{SB} \\
\midrule[0.8pt]
Crypto Primitives* & \checkmark & \checkmark & \checkmark & \checkmark & \checkmark & \checkmark & \checkmark \\
Secure Storage  & & \checkmark & \checkmark & \checkmark & \checkmark & & \checkmark \\
Remote Attestation & \checkmark & & \checkmark & & \checkmark & \checkmark & \\
AI Model Inference  & \checkmark & & &  & & & \\
Sec. Comm./I/O H.* & & \checkmark & \checkmark & & \checkmark & \checkmark & \\
Access Control  & & \checkmark & \checkmark & & \checkmark & \checkmark & \checkmark \\
\bottomrule[1.5pt]
\end{tabular}
\begin{tablenotes}
\footnotesize
\item[*] \textbf{AI}: Artificial Intelligence; \textbf{IoT}: Internet of Things; \textbf{BC}: Blockchain; \textbf{DB}: Database; \textbf{PP}: Privacy-Preserving; \textbf{NC}: Network Communication; \textbf{SB}: Secure Boot; \textbf{Sec. Comm./I/O H.*}: Secure Communication / I/O Handling; \textbf{Crypto Primitives*}: Cryptographic Primitives.
\end{tablenotes}
\end{threeparttable}
\end{table}

To explore how these functionalities manifest across different \textit{application domains}, we also constructed a cross-domain mapping (Table~\ref{tab:function-cross-domain}). The results show that some functionalities are nearly universal, appearing across all domains to build foundational security guarantees, while others remain highly domain-specific. Foundational primitives such as \textit{cryptographic operations} and \textit{secure storage}, which account for 68.2\% and 43.6\% of all projects respectively according to Table~\ref{tab:functionalities}, appear across almost all domains and serve as the core trust anchors of TEE-based systems. In contrast, advanced functionalities such as \textit{AI model inference} (14.0\%) and \textit{secure communication handling} (16.7\%) tend to cluster within specific contexts such as AI inference pipelines or networked edge devices, which is expected given their specialized requirements.

\vspace{0.1cm}
\noindent
\textbf{Key Insights.}
In addition to the distribution of categories, we further derive key insights from analyzing the detailed implementation of each TEE functionality.

\vspace{0.08cm}
\noindent
\textit{Pattern 1: Modularity Integration.}
TEE usage in real-world systems is typically modular. Although recent research prototypes~\cite{li2023survey,zheng2021survey, paju2023sok} have made running entire applications inside TEEs more practical, developers seldom migrate full software stacks. Instead, they encapsulate only critical components such as cryptographic primitives that require strong confidentiality guarantees. This modular design reflects a pragmatic understanding of TEE limitations, including constrained memory and context switching overhead~\cite{gentilal2017trustzone,paju2023sok,ma2024bitdb}, balancing protection requirements against performance costs. It also highlights a divergence between design intent and real-world practice. While TEEs were originally envisioned as general purpose trusted containers~\cite{mcgillion2015open,arnautov2016scone}, developers often employ them as \textit{lightweight trust anchors}~\cite{vuillermoz2022analysis} within larger system architectures to safeguard specific assets or operations. For example, a privacy preserving analytics service may perform data aggregation outside the TEE while confining attestation within it, which simplifies development but fragments the overall trust boundary, potentially introducing new attack surfaces as data flows between trusted and untrusted components.

\vspace{0.08cm}
\noindent
\textit{Pattern 2: Reimplementation and Bypassing of Standard TEE APIs.}
\label{sec:pattern1}
Another key finding is that many TEE projects reimplement cryptographic functionalities such as encryption and hashing algorithms instead of relying on official SDK-provided APIs. This trend is particularly evident in Intel SGX-based projects but is also observed in ARM TrustZone. Our quantitative analysis shows that in total 78 out of 241 projects (32.4\%) manually implement cryptographic operations within enclaves or TAs, including 53 out of 149 SGX projects (35.6\%) and 25 out of 92 TrustZone projects (27.1\%), \rev{such as AsusWRT Merlin~\cite{flashrouters_asus_merlin}, ppfl~\cite{mo2021ppfl} and mbedTLS~\cite{githubGitHubMbedTLSmbedtls}.} %These reimplementations commonly include symmetric encryption modes (e.g., AES-CTR~\cite{lee2022efficient}), hashing algorithms (e.g., SHA-256~\cite{gueron2011sha}), and asymmetric primitives for signing and key parsing~\cite{pointcheval2022asymmetric}.

Manual reimplementation can arise from various motivations. Through manual inspection of code comments, commit messages, and internal discussions, we qualitatively identified two main drivers: (1) The limited functionality or unclear documentation of official SDKs. For example, the Intel SGX SDK exposes only a subset of cryptographic primitives~\cite{intel_sgx_sdk}, while OP-TEE developers frequently report compatibility issues when integrating cryptographic APIs with modern toolchains~\cite{githubOPTEEoptee_os}. As a result, developers often import third-party cryptographic libraries such as mbedTLS~\cite{githubGitHubMbedTLSmbedtls} and OpenSSL~\cite{opensslOpenSSL}, or build lightweight wrappers around untrusted counterparts, thereby compromising the original isolation guarantees of TEEs; (2) The desire for greater flexibility and control over cryptographic implementations. Developers may choose to customize algorithms or port existing implementations to meet specific security or performance requirements that official SDKs may not support without extensive modification. For instance, developers frequently port OpenSSL~\cite{opensslOpenSSL} into enclaves to leverage its rich feature set and optimizations, as observed in projects such as \texttt{ppfl} and \texttt{darknetz}~\cite{mo2020darknetz,mo2021ppfl}.

While manual reimplementation can be functionally correct, it may inadvertently undermine the security guarantees that TEEs are designed to provide. First, reimplemented primitives can introduce logic errors or side-channel vulnerabilities that are absent in officially verified SDKs. Second, many projects omit critical initialization steps such as secure random generation, which weakens overall cryptographic robustness. Finally, divergence from standardized APIs reduces portability and maintainability across TEE platforms, hindering efforts to build reusable secure modules. Collectively, these findings reveal a persistent tension between the usability of TEE SDKs and the security of their real-world implementations.

An illustrative snippet from \texttt{ta\_sec\_key} in AsusWRT Merlin~\cite{flashrouters_asus_merlin}, a popular router firmware that secures private keys within TrustZone, is shown in Listing~\ref{lst:manual_aes_keyexp}. The snippet manually and correctly performs an AES-128 key expansion routine~\cite{lee2022efficient}, yet it exposes several weaknesses when executed inside a TEE. It relies on table lookups that are not constant time and may leak information through cache or timing side channels~\cite{van2018foreshadow}, even within enclaves or TAs. Moreover, the function lacks proper memory hygiene, leaving temporary buffers and round keys uncleared in memory~\cite{shu2017study}, and it omits secure random initialization, which is essential for maintaining cryptographic strength~\cite{egele2013empirical}.

Overall, this pattern reveals a gap between the design intent of TEEs and their practical use in real-world development. Since TEEs are often viewed as black boxes that inherently ensure security, developers may underestimate the complexity of implementing secure cryptographic functionalities within them. When encountering difficulties with SDK APIs, they often choose to reimplement components themselves to achieve greater flexibility. This gap highlights the need to improve the usability of TEE SDKs through richer reference implementations, clearer documentation, and automated analysis tools that can detect or warn against unsafe reimplementations. In addition, supporting modular and pluggable cryptographic backends could allow developers to retain flexibility while preserving the core security guarantees of TEEs~\cite{li2023survey}.

\begin{lstlisting}[
    style=customc,
    float=t!,
    caption={Manual AES Key Expansion in \texttt{ta\_sec\_key}~\cite{flashrouters_asus_merlin}},
    label=lst:manual_aes_keyexp
]
static void KeyExpansion(uint8_t* RoundKey, const uint8_t* Key) {
    for (int i = 0; i < 16; ++i)
        RoundKey[i] = Key[i];
    for (int i = 4; i < 44; ++i) {
        uint8_t temp[4];
        for (int j = 0; j < 4; ++j)
            temp[j] = RoundKey[(i - 1) * 4 + j];
        if (i % 4 == 0) {
            uint8_t k = temp[0];
            temp[0] = sbox[temp[1]]; temp[1] = sbox[temp[2]];
            temp[2] = sbox[temp[3]]; temp[3] = sbox[k];
            temp[0] ^= Rcon[i / 4];
        }
        for (int j = 0; j < 4; ++j)
            RoundKey[i * 4 + j] = RoundKey[(i - 4) * 4 + j] ^ temp[j];
    }
}
\end{lstlisting}

\vspace{0.08cm}
\noindent{\it Pattern 3: AI Model Migration Exhibits Clear Design Intent.}
\label{sec:pattern2}
A notable subset of collected TEE projects focuses on enabling AI inference within TEEs. While early implementations were mostly proof-of-concept demos, recent projects reveal more mature and deliberate design choices. Unlike traditional security applications, which often vary widely in structure, these AI-related projects exhibit consistent architectural or design patterns that reflect developers' collective efforts to balance performance, executability, and privacy under TEE constraints.
Across both Intel SGX and ARM TrustZone projects, we identified three dominant design strategies for migrating neural models into TEEs:
\begin{itemize}[leftmargin=*]
\item \textbf{Modularized Multi-Model Deployment.} Projects such as \texttt{DarkneTZ} isolate multiple model variants (e.g., TinyYOLO~\cite{khokhlov2020tiny}, MobileNet~\cite{sinha2019thin}) into separate enclave modules or TAs, enabling dynamic selection or replacement without redeploying the entire enclave.
\item \textbf{Simplified Single-Model Enclaves.} Lightweight CNNs such as MiniVGG~\cite{pyimagesearchMiniVGGNetGoing} are embedded as single-purpose inference TAs by projects like \texttt{ShadowNet}~\cite{sun2023shadownet}. These models are typically statically linked, quantized, or reduced in depth to fit within the limited enclave memory space.
\item \textbf{Compiler-Aided Optimization.} Projects like \texttt{Enigma}~\cite{li2022enigma} apply ahead-of-time (AOT) simplification~\cite{thom2018survey} and runtime code generation~\cite{chen2018tvm} to produce enclave-optimized inference binaries during training.
\end{itemize}

Our dataset contains 28 AI-in-TEE projects (out of 241), which we further categorized into 11 modularized designs, 9 simplified static deployments, and 8 compiler-aided approaches. Each strategy represents a distinct trade-off. Modularization enhances code reuse and supports multi-model scenarios but increases inter-TA communication complexity. Simplification reduces the enclave footprint and attack surface but may degrade inference accuracy. Compiler-aided optimization improves performance and binary compactness but requires nontrivial tooling and preprocessing during training.

Although these strategies differ in implementation, they share a common goal of adapting AI workloads to the unique constraints of TEEs, such as limited memory (e.g., the 128 MB limit of Intel SGX 1.0~\cite{intel_sgx_sdk}) and the performance overhead introduced by enclave transitions~\cite{paju2023sok}. By analyzing the source code and documentation of these projects, we observed that developers consciously select strategies based on their use cases and deployment environments, focusing on simplicity and optimization to ensure that AI models can be executed efficiently within TEEs without excessive overhead or complexity. This trend reflects a growing maturity in confidential AI development and suggests that TEE-based AI deployment is evolving from ad hoc experimentation toward more systematic engineering practices. It also motivates further research into specialized TEE support for AI workloads, such as reducing context-switch costs~\cite{li2023survey} and enabling larger memory capacity and secure model updates~\cite{yuan2024hypertheft}.

% \begin{tcolorbox}[
%   colback=gray!10!white,
%   colframe=frameblue,
%   title=RQ2 Summary - Observed Patterns in TEE Development,
%   boxsep=4pt,          % 减少文本与边框的距离（默认约为 4pt）
%   left=2pt, right=2pt, % 控制左右内边距
%   top=2pt, bottom=2pt, % 控制上下内边距
%   fontupper=\small     % 调整文字大小
% ]
% Our analysis of 241 TEE projects shows that developers most often implement \textit{cryptographic primitives} (68.2\%) and \textit{secure storage} (43.6\%) inside TEEs. We also identify three recurring patterns: (1) modularizing TEEs to secure critical components, (2) reimplementing cryptographic logic for flexibility (observed in 32.3\% of projects), and (3) adopting deliberate strategies for AI model deployment.
% \end{tcolorbox}

\ans{\textbf{RQ2 Summary:} Developers most frequently implement \textit{cryptographic primitives} (68.2\%), with 32.4\% of projects reimplementing SDK cryptographic APIs to improve flexibility or usability. Developers also tend to modularize their applications to protect critical components and adopt deliberate strategies for AI model deployment within TEEs.}

% Our analysis of 241 TEE projects reveals that developers most frequently implement \textit{cryptographic primitives} (68.2\%) and \textit{secure storage} (43.6\%) inside TEEs. We further identify recurring implementation patterns: developers often (1) modularize TEEs to protect only critical components, (2) reimplement cryptographic logic for flexibility at the cost of safety (observed in 32.3\% of projects), and (3) employ deliberate strategies for deploying AI models within TEEs.

\subsection{RQ3: What insecure practices and pitfalls are prevalent in real-world TEE projects?}
\label{sec:findings-misuse}
While TEEs provide hardware-enforced isolation, their security guarantees can still be undermined by insecure coding practices and configuration mistakes. Prior research has identified several common pitfalls in TEE usage, such as missing input validation~\cite{ma2025diting}, side-channel vulnerabilities~\cite{van2018foreshadow}, and flawed key management~\cite{liu2022extending}. However, these analyses typically focus on theoretical vulnerabilities or isolated case studies, and the prevalence as well as diversity of insecure practices in real-world TEE applications remain largely unexplored.

\vspace{0.1cm}
\noindent{\bf Categorization Process.}
To bridge this gap, we systematically examined the source code of 241 TEE projects to uncover recurring insecure or discouraged practices that deviate from established secure coding guidelines. Rather than conducting exhaustive vulnerability assessments, our goal is to identify commonly recurring patterns, assess their potential impact, and derive actionable recommendations for improving TEE security in practice. Because no mature static or dynamic analysis tools currently exist for systematically auditing TEE code, existing methods such as Ma et al.~\cite{ma2025diting} can only detect a limited number of isolated cases and lack perfect accuracy. Therefore, our investigation relies primarily on manual code inspection, supported by the customized static analysis introduced in Section~\ref{sec:methodology}. Specifically, we again used our static analysis scripts that apply keyword matching, prefix detection, and directory pattern recognition to identify (i) the presence of TEE-specific API calls like \texttt{TEE\_} and \texttt{sgx\_}, and (ii) the location of source files within TEE-specific directories (e.g., \texttt{ta/}, \texttt{enclave/}, \texttt{sgx/}). Using the matched code context of these TEE-specific artifacts, we then manually analyzed all functions and tagged those that exhibited insecure practices.

To ensure both completeness (minimizing false negatives, i.e., missed insecure cases) and accuracy (reducing false positives, i.e., benign examples mistakenly flagged as insecure), we designed a three-stage labeling protocol. First, two authors independently inspected each repository by following Intel SGX and OP-TEE documentation, official SDK coding guidelines, and prior TEE vulnerability taxonomies such as Ma et al.~\cite{ma2025diting}. Second, the reviewers conducted calibration sessions on a pilot subset to refine the taxonomy and ensure consistent interpretation of insecure patterns. Third, a third author performed cross-validation, focusing on ambiguous cases to confirm that each identified instance represented a genuine security concern rather than benign code. This categorization process took approximately 100 hours in total, which was longer than the time required for the other two RQs.

% This process, combining human expertise and rule-based static matching, compensates for the current lack of TEE-specific analysis tools and ensures high inter-reviewer consistency. It allows us to systematically capture recurring insecure behaviors that automated approaches would likely miss due to the absence of mature enclave-aware analyzers.

\vspace{0.1cm}
\noindent{\bf Categorization Results.}
Based on this process, we identified five recurring categories of insecure practices prevalent in real-world TEE projects, as introduced below.

\begin{itemize}[leftmargin=*]
  \item \textbf{Sensitive Data Leaks in Logs:} Developers often print cryptographic materials (e.g., plaintexts, salts, or digests) using debugging functions such as \texttt{printf}, inadvertently exposing secrets to untrusted environments.
  \item \textbf{Hardcoded Secrets or Initialization Vectors (IVs):} Cryptographic keys, IVs, or authentication tokens are embedded directly in enclave source code, with no randomness, rotation, or update mechanism.
  \item \textbf{Missing Input Validation:} Memory buffers, key lengths, and message structures are used without proper bounds checking, leading to potential buffer overflows or parsing errors~\cite{ma2025diting}.
  \item \textbf{Lack of Caller or Attestation Checks:} Some enclaves process requests from any caller without verifying the client's identity or establishing a secure attestation channel.
  \item \textbf{Hardcoded Paths or Constants:} File system paths, configuration values, and identifiers are hardcoded, reducing portability and enabling privilege escalation attacks.
\end{itemize}

Among the 241 TEE-based projects we examined, 61 projects (approximately 25.3\%) exhibited at least one insecure practice. Specifically, we identified 19 projects that leaked secrets through logging mechanisms, 25 that used hardcoded cryptographic keys or static IVs, 11 that lacked input validation, and 6 that failed to perform adequate identity or attestation checks. Notably, multiple security issues often co-occurred within the same project, significantly amplifying the overall attack surface. While some of these issues may originate from rapid prototyping or testing, our findings indicate that they are frequently carried over into production code, which suggests that current SDKs and development frameworks should provide clearer guidance and built-in safeguards to prevent such insecure implementations.

\vspace{0.1cm}
\noindent{\bf Key Insights.}
The prevalence of these insecure patterns reveals a critical gap between the security guarantees envisioned by TEE designers and the realities of software engineering practice. Developers often underestimate the sensitivity of in-enclave operations or place excessive trust in the assumed inviolability of TEE boundaries. Through manual inspection of code and internal discussions, we qualitatively argue that this misconception is likely rooted in several factors:
(1) a lack of formal training and awareness of TEE-specific secure coding practices;
(2) limited security-focused documentation and examples in official SDKs; and
(3) the inherent complexity of TEE development, which lacks mature tooling and automated analysis support compared to traditional software development, as also highlighted by prior studies~\cite{li2023survey,ma2025diting}.
These findings collectively underscore the urgent need for better developer education, improved SDK usability, and dedicated tooling support to help developers identify and avoid common pitfalls when building TEE-based applications.

To illustrate these insecure practices in more detail, we present two representative code snippets that exemplify common pitfalls. Each example highlights a specific insecure practice, explains how it occurs and its implications, and discusses how it undermines the security guarantees of TEEs.

\vspace{0.1cm}
\noindent{\it Fixed Initialization Vector Usage.}
The first case concerns improper cryptographic parameter handling. AES-GCM encryption~\cite{lee2022efficient} requires a unique IV for every encryption to ensure semantic security. However, as shown below, the \texttt{BI-SGX} project initializes the IV as an all-zero array, violating this requirement. This design flaw likely stems from a misunderstanding of IV requirements or from developers reusing initialization templates for convenience. It allows identical plaintext blocks encrypted under the same key and IV to produce identical ciphertexts, enabling attackers to infer data patterns or recover plaintexts~\cite{bock2016nonce}. Critically, this vulnerability arises even if the encryption key remains securely sealed within the TEE, demonstrating that hardware isolation alone cannot compensate for weak cryptographic practice.
\begin{lstlisting}[
    style=customc,
    numbers=none,
    xleftmargin=0em,
    morekeywords={sgx_rijndael128GCM_encrypt},
    float=h,
    caption={Hardcoded IV in AES-GCM (from \texttt{BI-SGX})},
    label=lst:fixed_iv
]
uint8_t fixed_iv[12] = {0};
sgx_rijndael128GCM_encrypt(&aes_key, plaintext, pt_len, ciphertext, fixed_iv, sizeof(fixed_iv), NULL, 0, &mac);
\end{lstlisting}

\noindent{\it Missing Parameter Checks.}
The second case illustrates unsafe memory handling and a lack of input validation. In \texttt{ta\_sec\_key}, code within the TEE copies user-provided data directly into a private key buffer and subsequently parses it as a key without verifying its size or structure. Such unchecked operations likely stem from developers assuming that input data is already sanitized by the untrusted world or from prioritizing simplicity and performance over security. This practice risks out-of-bounds memory access or buffer overflows, opening the possibility for malformed key injection that can lead to undefined behavior or privilege escalation inside the enclave. Because enclaves frequently process untrusted inputs originating from the normal world, robust input validation is critical yet often overlooked in practice.
\begin{lstlisting}[
    style=customc,
    numbers=none,
    xleftmargin=0em,
    morekeywords={TEE_MemMove},
    float=h,
    caption={Unchecked input before parsing DER keys (from \texttt{ta\_sec\_key})},
    label=lst:no_input_validation
]
TEE_MemMove(private_key_buffer, params[0].memref.buffer, params[0].memref.size); // No input bounds check
parse_der_private_key(private_key_buffer); // Unsafe use
\end{lstlisting}

\noindent{\it Leaking Secrets in Debug Output.}
This case demonstrates how poor logging practices can inadvertently nullify the confidentiality guarantees of TEEs. The code snippet from \texttt{comcast\_crypto\_gp\_ta} logs both cryptographic keys and intermediate digest values directly through the enclave's debug output. Such practices often arise from developers enabling verbose logging for debugging or performance tracing during testing, without properly disabling it in production builds. While convenient for development, they expose sensitive data to the untrusted operating system, effectively bypassing enclave isolation. Once printed, these secrets may be accessible via system logs or debugging tools, leaving persistent forensic traces even after enclave termination. This highlights a recurring misconception among developers that operations executed within a TEE are inherently safe regardless of output channels.
\begin{lstlisting}[
    style=customc,
    numbers=none,
    xleftmargin=0em,
    morekeywords={IMSG},
    float=h,
    caption={Logging cryptographic materials (from \texttt{comcast\_crypto\_gp\_ta})},
    label=lst:log_leak
]
IMSG("Key: %s", key_buf);       // Sensitive material
IMSG("Digest: %s", digest_result);
\end{lstlisting}

% These insecure practices undermine core security guarantees expected from TEE-based systems. For instance, the reuse of initialization vectors (IVs) in encryption schemes like AES-GCM is particularly dangerous, as it enables plaintext recovery and forgery attacks~\cite{bock2016nonce}. Additionally, the absence of proper parameter validation can lead to memory corruption bugs such as out-of-bounds access or crashes~\cite{ma2025diting}. Information leakage through debug logs is also a recurring issue~\cite{zhou2020mobilogleak}, where developers inadvertently bypass TEE isolation by exposing secrets to untrusted environments.

% \begin{tcolorbox}[colback=gray!10!white, colframe=frameblue, title=RQ3 Summary - Insecure Practices in TEE Development,  boxsep=4pt,          % 减少文本与边框的距离（默认约为 4pt）
%   left=2pt, right=2pt, % 控制左右内边距
%   top=2pt, bottom=2pt, % 控制上下内边距
%   fontupper=\small     % 调整文字大小
% ]
% Over a quarter of TEE projects (25.3\%) exhibit insecure practices such as secret leakage in logs, hardcoded cryptographic values, and missing input validation. These issues undermine the intended security guarantees of TEEs and underscore the need for clearer development guidelines, safer SDK defaults, and automated tools for detecting insecure patterns.
% \end{tcolorbox}

\ans{\textbf{RQ3 Summary:} Over a quarter of TEE projects (25.3\%) exhibit insecure practices such as secret leakage in logs and missing input validation. These weaknesses undermine the intended security guarantees of TEEs and highlight the need for clearer development guidelines and automated tools to detect and prevent insecure patterns.}

\section{Discussion}\label{sec:discussion}

Our empirical analysis reveals that while TEEs provide strong theoretical guarantees, their practical adoption often departs from ideal security models. We discuss these findings through three complementary perspectives: security, usability, and design, and conclude with directions for improvement.

\vspace{0.1cm}
\noindent
\textbf{Security Viewpoint: Practical Deviations from Ideal Models.}
Many projects diverge from the secure programming principles envisioned by TEEs. Common problems include logging sensitive data such as keys or authentication tokens using debug macros like \texttt{printf}, skipping attestation or verification steps, and hardcoding secrets in trusted code. We also find frequent misuse of cryptographic APIs, such as reusingIVrs and re-implementing primitives~\cite{egele2013empirical,nadi2016jumping}.
Similar patterns have been reported in other security contexts like cryptographic libraries~\cite{fahl2012eve,acar2016you}, but their impact is more severe in TEEs because enclave code operates under high trust assumptions. Once a vulnerability is introduced, it undermines the very isolation guarantees the hardware is meant to enforce~\cite{chen2018sgxpectre}. These deviations often stem from a combination of insufficient security awareness and practical barriers. Many developers rely on reused testing code or incomplete examples, and some face difficulties using complex attestation or cryptographic APIs, which leads to shortcuts that compromise security.

\vspace{0.1cm}
\noindent
\textbf{Usability Viewpoint: Re-Implementation as a Signal of Friction.}
Frequent reimplementation of core SDK functions such as \texttt{TEE\_AllocateOperation()} and custom handling of sessions or memory suggest that current TEE SDKs present substantial usability challenges. Developers often modify or bypass provided abstractions to match real-world workflows, revealing a gap between SDK design and actual development needs~\cite{wijayarathna2019using}.
Such behaviors are not necessarily misuse but signals of friction. Developers adapt APIs when abstractions are too rigid, documentation lacks clarity, or interfaces remain overly hardware-centric. To make TEEs more approachable, SDKs should emphasize simplicity, clear examples, and safer defaults that guide developers toward correct usage rather than penalize experimentation.

\vspace{0.1cm}
\noindent
\textbf{Design Viewpoint: Emerging Good Practices.}
Despite the issues discussed above, we also observe promising design patterns, particularly in AI-related projects. Developers increasingly partition workloads into modular components, deploy lightweight neural models that fit within enclave memory, and apply compiler-assisted transformations to generate optimized enclave binaries~\cite{mo2020darknetz,li2022enigma}. These design choices reflect a growing understanding of TEE constraints. Rather than treating TEEs solely as secure vaults for protecting keys, developers are integrating them as essential components of system architectures to execute more complex tasks securely. This shift demonstrates a maturation in TEE usage, where developers achieve a more effective balance between performance, usability, and security, and it also suggests rising pressure in practice to deliver secure systems. In this context, implementations that encapsulate entire applications within TEEs can be viewed as pragmatic solutions to avoid partitioning.
% , even though our dataset contains no direct evidence of such cases.

\vspace{0.1cm}
\noindent
\textbf{Future Directions.}
Our findings point to several directions for improvement. First, SDK designers should enhance developer support through clearer documentation, simplified cryptographic interfaces, and secure code examples and project templates for common TEE development scenarios. Second, tool developers can build program analyzers~\cite{ma2025diting} to detect TEE-specific insecure API usage, and other recurring issues. Finally, researchers can design developer-friendly frameworks that make TEEs easier to use without sacrificing performance or security, thereby narrowing the gap between TEE's theoretical guarantees and their practical deployment.
In summary, the effectiveness of TEEs depends not only on hardware isolation but also on how developers understand and adopt them. Bridging the gap between idealized security assumptions and practical development remains crucial, and our study provides empirical evidence to support progress toward this goal.

% Threats to Validity
\section{Threats to Validity}

\vspace{0.1cm}
\noindent
\textbf{External Validity.}
Our dataset focuses on open-source projects using Intel SGX and ARM TrustZone (via OP-TEE), the two most widely-adopted TEEs. Other platforms such as RISC-V Keystone~\cite{lee2020keystone} were excluded due to limited adoption in open-source ecosystems. While our findings may not fully generalize to proprietary or industrial deployments, we mitigated this threat by including both production-grade and prototype projects across diverse domains to ensure representativeness.

\vspace{0.1cm}
\noindent
\textbf{Construct Validity.}
To ensure the maturity of the analyzed repositories and that they reflect genuine developer practice, we required each project to have at least 10 stars and verified the presence of TEE-related artifacts, such as API calls. \rev{Although our automated screening is grounded in official documentation and SDK artifacts, it may still miss projects that invoke TEE functionality through non-standard wrappers, potentially causing some TEE-specific code to be overlooked. To mitigate this issue, we incorporate a broad set of indicators to assess whether TEE implementations are actually present, including not only TEE-specific API calls, but also platform-specific build or configuration files, as well as directory and file naming conventions. This enables us to retrieve a set of potentially TEE-related projects as comprehensively as possible, which are subsequently verified through manual inspection (see Section~\ref{sec:methodology}).}
% \textbf{Construct Validity.}
% Some analyzed repositories are research prototypes rather than production systems, but they still reveal developer intent and SDK usability issues. To improve validity, we excluded trivial examples, required at least 10 GitHub stars to ensure maturity and relevance, and verified TEE usage by inspecting TEE artifacts such as API calls, and build files.\rev{Additionally, our initial screening relied on regex-based keyword matching derived from official documentation. While we iteratively refined these keywords to maximize coverage, there remains a possibility that projects using highly non-standard wrappers or obfuscated API calls were missed (false negatives), which is a common limitation in keyword-based empirical studies.}
% Some analyzed repositories are research prototypes or demos rather than production systems. Nonetheless, they still reveal developer intent and SDK usability challenges. We excluded trivial scaffolds or non-functional examples and retained only repositories with substantive enclave or TA implementations, verified through repository metadata such as stars and commit activity.

\vspace{0.1cm}
\noindent
\textbf{Internal Validity.}
Our manual analysis may introduce subjectivity or misclassification. To minimize this risk, we cross-checked ambiguous cases and refined the tags iteratively among multiple authors. In addition, since the data was collected between January 2011 and May 2025, our findings reflect the state of TEE practices during this period and may evolve as newer SDKs and hardware releases emerge.

% Conclusion
\section{Conclusion and Future Work}
This paper presents the first large-scale empirical study of Trusted Execution Environments in the wild, collecting and analyzing 241 open-source Intel SGX and ARM TrustZone projects that were created or updated between 2011 and 2025. Our analysis shows that TEEs are most commonly used for cryptographic primitives (68.2\%), secure storage (43.6\%), and remote attestation (28.7\%), while AI model protection has recently emerged as a rapidly growing application domain (12\%). However, 32.4\% of the projects reimplement cryptographic functions instead of using official SDK APIs, and 25.3\% contain insecure coding practices such as hardcoded secrets, missing input validation, or secret leakage through logs. These findings reveal a significant gap between the theoretical isolation guarantees of TEEs and their implementation in real-world software. Our study further discusses implications for TEE SDK designers, secure development practices, and future research directions to address this gap.
In future, we plan to develop effective repair techniques to automatically fix the misuses revealed in TEE applications.
% In future work, we plan to conduct surveys and interviews to understand TEE developers' design decisions and challenges. We also intend to perform a qualitative analysis of open-source issues and pull requests related to TEE usage to examine how security and usability concerns evolve over time. Additionally, we will explore the discussions surrounding the use of TEEs for developing AI systems in online forums and communities.

\vspace{0.1cm}
\noindent

\section*{Data Availability}
Our curated dataset and code for analysis are available at: \url{https://figshare.com/s/ae4ed260857cfd7df505}.

\section*{Acknowledgement}
This research / project is supported by the National Research Foundation, Singapore, and the Cyber Security Agency of Singapore under its National Cybersecurity R\&D Programme (Proposal ID: NCR25-DeSCEmT-SMU). Any opinions, findings and conclusions or recommendations expressed in this material are those of the author(s) and do not reflect the views of the National Research Foundation, Singapore, and the Cyber Security Agency of Singapore.

% % 参考文献前
\IEEEtriggeratref{52}                 % N 是触发分栏的位置（参考文献的序号）
% % \IEEEtriggercmd{\enlargethispage{-5in}}  % 需要时再开，用来微调最后一页高度
\bibliographystyle{IEEEtran}
\bibliography{references}

@inproceedings{paju2023sok,
  title={Sok: A systematic review of tee usage for developing trusted applications},
  author={Paju, Arttu and Javed, Muhammad Owais and Nurmi, Juha and Savim{\"a}ki, Juha and McGillion, Brian and Brumley, Billy Bob},
  booktitle={Proceedings of the 18th International Conference on Availability, Reliability and Security},
  pages={1--15},
  year={2023}
}

@article{costan2016intel,
  title={Intel SGX explained},
  author={Costan, Victor and Devadas, Srinivas},
  journal={Cryptology ePrint Archive},
  year={2016}
}

@inproceedings{mcgillion2015open,
  title={Open-TEE--an open virtual trusted execution environment},
  author={McGillion, Brian and Dettenborn, Tanel and Nyman, Thomas and Asokan, N},
  booktitle={{Trustcom/BigDataSE/ISPA}},
  volume={1},
  pages={400--407},
  year={2015},
  organization={IEEE}
}

@inproceedings{mo2020darknetz,
  title={{Darknetz}: towards model privacy at the edge using trusted execution environments},
  author={Mo, Fan and Shamsabadi, Ali Shahin and Katevas, Kleomenis and Demetriou, Soteris and Leontiadis, Ilias and Cavallaro, Andrea and Haddadi, Hamed},
  booktitle={Proceedings of the 18th International Conference on Mobile Systems, Applications, and Services},
  pages={161--174},
  year={2020}
}

@inproceedings{gentilal2017trustzone,
  title={TrustZone-backed bitcoin wallet},
  author={Gentilal, Miraje and Martins, Paulo and Sousa, Leonel},
  booktitle={Proceedings of the Fourth Workshop on Cryptography and Security in Computing Systems},
  pages={25--28},
  year={2017}
}

@inproceedings{kim2019shieldstore,
  title={{Shieldstore}: Shielded in-memory key-value storage with {SGX}},
  author={Kim, Taehoon and Park, Joongun and Woo, Jaewook and Jeon, Seungheun and Huh, Jaehyuk},
  booktitle={Proceedings of the Fourteenth EuroSys Conference},
  pages={1--15},
  year={2019}
}

@article{yang2023vedb,
  title={{VeDB}: A software and hardware enabled trusted relational database},
  author={Yang, Xinying and Zhang, Ruide and Yue, Cong and Liu, Yang and Ooi, Beng Chin and Gao, Qun and Zhang, Yuan and Yang, Hao},
  journal={Proceedings of the ACM on Management of Data},
  volume={1},
  number={2},
  pages={1--27},
  year={2023},
  publisher={ACM}
}

@manual{intel_sgx_sdk,
  title        = {{Intel SGX SDK Developer Reference}},
  author       = {{Intel Corporation}},
  year         = {2023},
  url          = {https://download.01.org/intel-sgx/latest/linux/docs/},
  note         = {Accessed: 2025-09-30}
}

@manual{optee_docs,
  title        = {{OP-TEE Documentation}},
  author       = {{TrustedFirmware Project}},
  year         = {2023},
  url          = {https://optee.readthedocs.io/},
  note         = {Accessed: 2025-09-30}
}

@article{zheng2021survey,
  title={A survey of Intel {SGX} and its applications},
  author={Zheng, Wei and Wu, Ying and Wu, Xiaoxue and Feng, Chen and Sui, Yulei and Luo, Xiapu and Zhou, Yajin},
  journal={Frontiers Comput. Sci.},
  volume={15},
  number={3},
  pages={153808},
  year={2021},
  publisher={Springer}
}

@inproceedings{shinde2017panoply,
  title={Panoply: Low-TCB Linux Applications With {SGX} Enclaves.},
  author={Shinde, Shweta and Le Tien, Dat and Tople, Shruti and Saxena, Prateek},
  booktitle={{NDSS}},
  year={2017}
}

@inproceedings{tsai2017graphene,
  title={{Graphene-SGX}: A practical library {OS} for unmodified applications on {SGX}},
  author={Tsai, Chia-Che and Porter, Donald E and Vij, Mona},
  booktitle={{USENIX} {ATC}},
  pages={645--658},
  year={2017}
}

@inproceedings{arnautov2016scone,
  title={{SCONE}: Secure linux containers with intel {SGX}},
  author={Arnautov, Sergei and Trach, Bohdan and Gregor, Franz and Knauth, Thomas and Martin, Andre and Priebe, Christian and Lind, Joshua and Muthukumaran, Divya and O'keeffe, Dan and Stillwell, Mark L and others},
  booktitle={{OSDI}},
  pages={689--703},
  year={2016}
}

@inproceedings{lee2020keystone,
  title={{Keystone}: An open framework for architecting trusted execution environments},
  author={Lee, Dayeol and Kohlbrenner, David and Shinde, Shweta and Asanovi{\'c}, Krste and Song, Dawn},
  booktitle={Proceedings of the Fifteenth European Conference on Computer Systems},
  pages={1--16},
  year={2020}
}

@inproceedings{mo2021ppfl,
  title={{PPFL}: Privacy-preserving federated learning with trusted execution environments},
  author={Mo, Fan and Haddadi, Hamed and Katevas, Kleomenis and Marin, Eduard and Perino, Diego and Kourtellis, Nicolas},
  booktitle={Proceedings of the 19th annual international conference on mobile systems, applications, and services},
  pages={94--108},
  year={2021}
}

@inproceedings{yuan2024hypertheft,
  title={{Hypertheft}: Thieving model weights from {TEE}-shielded neural networks via ciphertext side channels},
  author={Yuan, Yuanyuan and Liu, Zhibo and Deng, Sen and Chen, Yanzuo and Wang, Shuai and Zhang, Yinqian and Su, Zhendong},
  booktitle={{CCS}},
  pages={4346--4360},
  year={2024}
}

@article{han2025autotee,
  title={{AutoTEE}: Automated Migration and Protection of Programs in Trusted Execution Environments},
  author={Han, Ruidong and Yang, Zhou and Ma, Chengyan and Liu, Ye and Niu, Yuqing and Ma, Siqi and Gao, Debin and Lo, David},
  journal={arXiv preprint arXiv:2502.13379},
  year={2025}
}

@inproceedings{acar2016you,
  title={You get where you're looking for: The impact of information sources on code security},
  author={Acar, Yasemin and Backes, Michael and Fahl, Sascha and Kim, Doowon and Mazurek, Michelle L and Stransky, Christian},
  booktitle={{S\&P}},
  pages={289--305},
  year={2016},
  organization={IEEE}
}

@inproceedings{egele2013empirical,
  title={An empirical study of cryptographic misuse in android applications},
  author={Egele, Manuel and Brumley, David and Fratantonio, Yanick and Kruegel, Christopher},
  booktitle={{CCS}},
  pages={73--84},
  year={2013}
}

@article{wijayarathna2019using,
  title={Using cognitive dimensions to evaluate the usability of security APIs: An empirical investigation},
  author={Wijayarathna, Chamila and Arachchilage, Nalin Asanka Gamagedara},
  journal={Information and Software Technology},
  volume={115},
  pages={5--19},
  year={2019},
  publisher={Elsevier}
}

@inproceedings{fahl2012eve,
  title={Why Eve and Mallory love Android: An analysis of Android {SSL} (in) security},
  author={Fahl, Sascha and Harbach, Marian and Muders, Thomas and Baumg{\"a}rtner, Lars and Freisleben, Bernd and Smith, Matthew},
  booktitle={{CCS}},
  pages={50--61},
  year={2012}
}

@inproceedings{nadi2016jumping,
  title={Jumping through hoops: Why do Java developers struggle with cryptography {APIs}?},
  author={Nadi, Sarah and Kr{\"u}ger, Stefan and Mezini, Mira and Bodden, Eric},
  booktitle={{ICSE}},
  pages={935--946},
  year={2016}
}

@inproceedings{meng2018secure,
  title={Secure coding practices in java: Challenges and vulnerabilities},
  author={Meng, Na and Nagy, Stefan and Yao, Danfeng and Zhuang, Wenjie and Argoty, Gustavo Arango},
  booktitle={{ICSE}},
  pages={372--383},
  year={2018}
}

@article{gorski2016towards,
  title={Towards the usability evaluation of security {APIs}.},
  author={Gorski, Peter Leo and Iacono, Luigi Lo},
  journal={HAISA},
  volume={10},
  pages={252--265},
  year={2016}
}

@inproceedings{bove2024large,
  title={A large-scale study on the prevalence and usage of tee-based features on android},
  author={Bove, Davide},
  booktitle={{ARES}},
  pages={1--11},
  publisher= {{ACM}},
  year={2024}
}

@inproceedings{cosentino2016findings,
  title={Findings from {GitHub}: methods, datasets and limitations},
  author={Cosentino, Valerio and Luis, Javier and Cabot, Jordi},
  booktitle={{MSR}},
  pages={137--141},
  publisher= {{ACM}},
  year={2016}
}

@inproceedings{humbatova2020taxonomy,
  title={Taxonomy of real faults in deep learning systems},
  author={Humbatova, Nargiz and Jahangirova, Gunel and Bavota, Gabriele and Riccio, Vincenzo and Stocco, Andrea and Tonella, Paolo},
  booktitle={Proceedings of the ACM/IEEE 42nd international conference on software engineering},
  pages={1110--1121},
  year={2020}
}

@inproceedings{cruzes2011recommended,
  title={Recommended steps for thematic synthesis in software engineering},
  author={Cruzes, Daniela S and Dyba, Tore},
  booktitle={{ESEM}},
  pages={275--284},
  year={2011},
  organization={IEEE}
}

@inproceedings{stol2016grounded,
  title={Grounded theory in software engineering research: a critical review and guidelines},
  author={Stol, Klaas-Jan and Ralph, Paul and Fitzgerald, Brian},
  booktitle={{ICSE}},
  pages={120--131},
  year={2016}
}

@article{chen2018sgxpectre,
  title={Sgxpectre attacks: Leaking enclave secrets via speculative execution},
  author={Chen, Guoxing and Chen, Sanchuan and Xiao, Yuan and Zhang, Yinqian and Lin, Zhiqiang and Lai, Ten H},
  journal={arXiv preprint arXiv:1802.09085},
  year={2018}
}

@inproceedings{van2018foreshadow,
  title={{Foreshadow}: Extracting the keys to the intel {SGX} kingdom with transient {Out-of-Order} execution},
  author={Van Bulck, Jo and Minkin, Marina and Weisse, Ofir and Genkin, Daniel and Kasikci, Baris and Piessens, Frank and Silberstein, Mark and Wenisch, Thomas F and Yarom, Yuval and Strackx, Raoul},
  booktitle={{USENIX Security}},
  pages={991--1008},
  year={2018}
}

@inproceedings{shu2017study,
  title={A study of security vulnerabilities on docker hub},
  author={Shu, Rui and Gu, Xiaohui and Enck, William},
  booktitle={{CODASPY}},
  pages={269--280},
  year={2017}
}

@article{li2023survey,
  title={A survey of secure computation using trusted execution environments},
  author={Li, Xiaoguo and Zhao, Bowen and Yang, Guomin and Xiang, Tao and Weng, Jian and Deng, Robert H},
  journal={arXiv preprint arXiv:2302.12150},
  year={2023}
}

@article{ma2025diting,
  title={{Diting}: A static analyzer for identifying bad partitioning issues in tee applications},
  author={Ma, Chengyan and Han, Ruidong and Shi, Jieke and Liu, Ye and Niu, Yuqing and Lu, Di and Tian, Chuang and Ma, Jianfeng and Gao, Debin and Lo, David},
  journal={arXiv preprint arXiv:2502.15281},
  year={2025}
}

@inproceedings{bock2016nonce,
  title={{Nonce-Disrespecting} adversaries: Practical forgery attacks on {GCM} in {TLS}},
  author={B{\"o}ck, Hanno and Zauner, Aaron and Devlin, Sean and Somorovsky, Juraj and Jovanovic, Philipp},
  booktitle={{WOOT}},
  year={2016}
}

@inproceedings{arfaoui2014trusted,
  title={Trusted execution environments: A look under the hood},
  author={Arfaoui, Ghada and Gharout, Said and Traor{\'e}, Jacques},
  booktitle={2nd IEEE international conference on mobile cloud computing, services, and Engineering},
  pages={259--266},
  year={2014},
  organization={IEEE}
}

@article{jauernig2020trusted,
  title={Trusted execution environments: properties, applications, and challenges},
  author={Jauernig, Patrick and Sadeghi, Ahmad-Reza and Stapf, Emmanuel},
  journal={{S\&P}},
  volume={18},
  number={2},
  pages={56--60},
  year={2020},
  publisher={IEEE}
}

@misc{globalplatformAnnualReport,
	author = {GlobalPlatform, a consensus-driven technical standards organization},
	title = {{A}nnual {R}eport 2024 of {G}lobal{P}latform },
	howpublished = {\url{https://globalplatform.org/resource-publication/annual-report-2024/}},
	year = {2024}
}

@misc{growthmarketreportsTrustedExecution,
	author = {Growth Market Reports},
	title = {{T}rusted {E}xecution {E}nvironment {M}arket {R}esearch {R}eport 2033},
	howpublished = {\url{https://growthmarketreports.com/report/trusted-execution-environment-market}},
	year = {2024}
}

@inproceedings{wang2020building,
  title={Building and maintaining a third-party library supply chain for productive and secure {SGX} enclave development},
  author={Wang, Pei and Ding, Yu and Sun, Mingshen and Wang, Huibo and Li, Tongxin and Zhou, Rundong and Chen, Zhaofeng and Jing, Yiming},
  booktitle={{ICSE-SEIP}},
  pages={100--109},
  publisher={{ACM}},
  year={2020}
}

@inproceedings{zheng2016trustpay,
  title={{TrustPAY}: Trusted mobile payment on security enhanced {ARM} {TrustZone} platforms},
  author={Zheng, Xianyi and Yang, Lulu and Ma, Jiangang and Shi, Gang and Meng, Dan},
  booktitle={{ISCC}},
  pages={456--462},
  year={2016},
  organization={IEEE}
}

@INPROCEEDINGS{10628729,
  author={Yagawa, Takashi and Teruya, Tadanori and Suzaki, Kuniyasu and Abe, Hirotake},
  booktitle={{EuroS\&PW}},
  title={Delegating Verification for Remote Attestation Using TEE},
  pages={186--192},
  publisher={{IEEE}},
  year={2024}
}

@inproceedings{li2019teev,
  title={{TEEv}: Virtualizing trusted execution environments on mobile platforms},
  author={Li, Wenhao and Xia, Yubin and Lu, Long and Chen, Haibo and Zang, Binyu},
  booktitle    = {{VEE}},
  pages        = {2--16},
  publisher    = {{ACM}},
  year         = {2019}
}

@inproceedings{cerdeira2020sok,
  title={{SoK}: Understanding the prevailing security vulnerabilities in trustzone-assisted tee systems},
  author={Cerdeira, David and Santos, Nuno and Fonseca, Pedro and Pinto, Sandro},
  booktitle={{S\&P}},
  pages={1416--1432},
  year={2020},
  organization={IEEE}
}

@misc{joy2024physicalsoftwarebasedfault,
      title={Physical and Software Based Fault Injection Attacks Against {TEEs} in Mobile Devices: A Systemisation of Knowledge},
      author={Aaron Joy and Ben Soh and Zhi Zhang and Sri Parameswaran and Darshana Jayasinghe},
      year={2024},
      eprint={2411.14878},
      archivePrefix={arXiv},
      url={https://arxiv.org/abs/2411.14878},
}

@inproceedings{borges2016understanding,
  title={Understanding the factors that impact the popularity of {GitHub} repositories},
  author={Borges, Hudson and Hora, Andre and Valente, Marco Tulio},
  booktitle={{ICSME}},
  pages={334--344},
  year={2016},
  organization={IEEE}
}

@article{liu2022extending,
  author       = {Chun{-}Chi Liu and
                  Hechuan Guo and
                  Minghui Xu and
                  Shengling Wang and
                  Dongxiao Yu and
                  Jiguo Yu and
                  Xiuzhen Cheng},
  title        = {Extending On-Chain Trust to Off-Chain - Trustworthy Blockchain Data
                  Collection Using Trusted Execution Environment {(TEE)}},
  journal      = {{IEEE} Trans. Computers},
  volume       = {71},
  number       = {12},
  pages        = {3268--3280},
  year         = {2022}
}

@article{viera2005understanding,
  title={Understanding interobserver agreement: the kappa statistic},
  author={Viera, Anthony J and Garrett, Joanne M and others},
  journal={Fam med},
  volume={37},
  number={5},
  pages={360--363},
  year={2005}
}

@inproceedings{begel2014analyze,
  title={Analyze this! 145 questions for data scientists in software engineering},
  author={Begel, Andrew and Zimmermann, Thomas},
  booktitle={{ICSE}},
  pages={12--23},
  year={2014}
}

@inproceedings{yang2023users,
  title={What do users ask in open-source AI repositories? An empirical study of {GitHub} issues},
  author={Yang, Zhou and Wang, Chenyu and Shi, Jieke and Hoang, Thong and Kochhar, Pavneet and Lu, Qinghua and Xing, Zhenchang and Lo, David},
  booktitle={{MSR}},
  pages={79--91},
  year={2023},
  organization={IEEE}
}

@article{zhang2023pacta,
  title={{PACTA}: An IoT data privacy regulation compliance scheme using {TEE} and blockchain},
  author={Zhang, Yongxin and Yang, Jiacheng and Lei, Hong and Bao, Zijian and Lu, Ning and Shi, Wenbo and Chen, Bangdao},
  journal={{IEEE} Internet Things J.},
  volume={11},
  number={5},
  pages={8882--8893},
  year={2023}
}

@article{istvan2020towards,
  title={Towards software-defined data protection: {GDPR} compliance at the storage layer is within reach},
  author={Istv{\'a}n, Zsolt and Ponnapalli, Soujanya and Chidambaram, Vijay},
  journal={arXiv preprint arXiv:2008.04936},
  year={2020}
}

@misc{opensslOpenSSL,
  title        = {{O}pen{S}SL {C}ryptography and {S}SL/TLS {T}oolkit},
  author       = {{OpenSSL Project}},
  year         = {2023},
  url          = {https://www.openssl.org/},
}

@phdthesis{vuillermoz2022analysis,
  title={Analysis of {TEE} technologies as trust anchors},
  author={Vuillermoz, Simone},
  year={2022},
  school={Politecnico di Torino}
}

@article{ma2024bitdb,
  title={{BiTDB}: Constructing a built-in {TEE} secure database for embedded systems},
  author={Ma, Chengyan and Lu, Di and Lv, Chaoyue and Xi, Ning and Jiang, Xiaohong and Shen, Yulong and Ma, Jianfeng},
  journal={{IEEE} Trans. Knowl. Data Eng.},
  volume={36},
  number={9},
  pages={4472--4485},
  year={2024}
}

@article{lee2022efficient,
  title={Efficient implementation of {AES-CTR} and {AES-ECB} on {GPUs} with applications for high-speed {FrodoKEM} and exhaustive key search},
  author={Lee, Wai-Kong and Seo, Hwa Jeong and Seo, Seog Chung and Hwang, Seong Oun},
  journal={{IEEE} Trans. Circuits Syst. {II} Express Briefs},
  volume={69},
  number={6},
  pages={2962--2966},
  year={2022}
}

@misc{githubGitHubMbedTLSmbedtls,
	author = {{M}bed-{T}{L}{S}},
	title = {{M}bed-{T}{L}{S}/mbedtls: {A}n open source, portable, easy to use, readable and flexible {T}{L}{S} library, and reference implementation of the {P}{S}{A} {C}ryptography {A}{P}{I}. {R}eleases are on a varying cadence, typically around 3 - 6 months between releases},
	howpublished = {\url{https://github.com/Mbed-TLS/mbedtls}},
	year = {2025},
}

@misc{githubOPTEEoptee_os,
	author = {OP-TEE},
	title = {{O}{P}-{T}{E}{E}/optee\_os},
	howpublished = {\url{https://github.com/OP-TEE/optee_os/issues}},
	note = {[Accessed 14-10-2025]}
}

@misc{flashrouters_asus_merlin,
  title        = {AsusWRT Merlin},
  author       = {{FlashRouters}},
  howpublished = {\url{https://www.flashrouters.com/router-basics/what-is-asus-merlin}},
  year         = 2025
}

@inproceedings{khokhlov2020tiny,
  title={{Tiny-YOLO} object detection supplemented with geometrical data},
  author={Khokhlov, Ivan and Davydenko, Egor and Osokin, Ilya and Ryakin, Ilya and Babaev, Azer and Litvinenko, Vladimir and Gorbachev, Roman},
  booktitle={{VTC}},
  pages={1--5},
  year={2020},
  organization={IEEE}
}

@inproceedings{sinha2019thin,
  title={Thin {MobileNet}: An Enhanced {MobileNet} Architecture},
  author={Sinha, Debjyoti and El-Sharkawy, Mohamed},
  booktitle={{UEMCON}},
  pages={0280--0285},
  year={2019},
  organization={IEEE}
}

@inproceedings{sun2023shadownet,
  title={Shadownet: A secure and efficient on-device model inference system for convolutional neural networks},
  author={Sun, Zhichuang and Sun, Ruimin and Liu, Changming and Chowdhury, Amrita Roy and Lu, Long and Jha, Somesh},
  booktitle={{S\&P}},
  pages={1596--1612},
  year={2023},
  organization={IEEE}
}

@misc{pyimagesearchMiniVGGNetGoing,
	author = {Adrian Rosebrock},
	title = {{M}ini{V}{G}{G}{N}et: {G}oing {D}eeper with {C}{N}{N}s - {P}y{I}mage{S}earch --- pyimagesearch.com},
	howpublished = {\url{https://pyimagesearch.com/2021/05/22/minivggnet-going-deeper-with-cnns/}},
	year = {2025},
}

@inproceedings{thom2018survey,
  title={A survey of ahead-of-time technologies in dynamic language environments},
  author={Thom, Mark and Dueck, Gerhard W and Kent, Kenneth and Maier, Daryl},
  booktitle={{CASCON}},
  pages={275--281},
  year={2018}
}

@inproceedings{li2022enigma,
  title={{ENIGMA}: Low-latency and privacy-preserving edge inference on heterogeneous neural network accelerators},
  author={Li, Qiushi and Ren, Ju and Pan, Xinglin and Zhou, Yuezhi and Zhang, Yaoxue},
  booktitle={{ICDCS}},
  pages={458--469},
  year={2022},
  organization={IEEE}
}

@inproceedings{chen2018tvm,
  title={{TVM}: An automated {End-to-End} optimizing compiler for deep learning},
  author={Chen, Tianqi and Moreau, Thierry and Jiang, Ziheng and Zheng, Lianmin and Yan, Eddie and Shen, Haichen and Cowan, Meghan and Wang, Leyuan and Hu, Yuwei and Ceze, Luis and others},
  booktitle={{OSDI}},
  pages={578--594},
  year={2018}
}

@inproceedings{galappaththi2024empirical,
  title={An empirical study of API misuses of data-centric libraries},
  author={Galappaththi, Akalanka and Nadi, Sarah and Treude, Christoph},
  booktitle={{ESEM}},
  pages={245--256},
  publisher={{ACM}},
  year={2024}
}

@inproceedings{arzt2015towards,
  title={Towards secure integration of cryptographic software},
  author={Arzt, Steven and Nadi, Sarah and Ali, Karim and Bodden, Eric and Erdweg, Sebastian and Mezini, Mira},
  booktitle={{Onward!}},
  pages={1--13},
  year={2015}
}

@article{mousavi2025detecting,
  title={Detecting misuse of security APIs: A systematic review},
  author={Mousavi, Zahra and Islam, Chadni and Babar, Muhammad Ali and Abuadbba, Alsharif and Moore, Kristen},
  journal={ACM Computing Surveys},
  volume={57},
  number={12},
  pages={1--39},
  year={2025}
}

@inproceedings{apostolopoulos1999transport,
  title={Transport Layer Security: How much does it really cost?},
  author={Apostolopoulos, George and Peris, Vinod and Saha, Debanjan},
  booktitle={{INFOCOM}},
  publisher={{IEEE}},
  year={1999},
}

@inproceedings{zhang2022teeslice,
  title={{TEESlice}: slicing {DNN} models for secure and efficient deployment},
  author={Zhang, Ziqi and Ng, Lucien KL and Liu, Bingyan and Cai, Yifeng and Li, Ding and Guo, Yao and Chen, Xiangqun},
  booktitle={{AISTA}},
  pages={1--8},
  year={2022}
}

@inproceedings{lo2023trustworthy,
  title={Trustworthy and synergistic artificial intelligence for software engineering: Vision and roadmaps},
  author={Lo, David},
  booktitle={{ICSE-FoSE}},
  pages={69--85},
  year={2023},
  organization={IEEE}
}

@article{liu2025towards,
  title={Towards Secure Program Partitioning for Smart Contracts with {LLM}'s In-Context Learning},
  author={Liu, Ye and Niu, Yuqing and Ma, Chengyan and Han, Ruidong and Ma, Wei and Li, Yi and Gao, Debin and Lo, David},
  journal={arXiv preprint arXiv:2502.14215},
  year={2025}
}

@article{lyu2025my,
  title={" My productivity is boosted, but..." Demystifying Users' Perception on AI Coding Assistants},
  author={Lyu, Yunbo and Yang, Zhou and Shi, Jieke and Chang, Jianming and Liu, Yue and Lo, David},
  journal={arXiv preprint arXiv:2508.12285},
  year={2025}
}

% \begin{thebibliography}{}
% \bibitem{b1} G. Eason, B. Noble, and I. N. Sneddon, ``On certain integrals of Lipschitz-Hankel type involving products of Bessel functions,'' Phil. Trans. Roy. Soc. London, vol. A247, pp. 529--551, April 1955.
% \bibitem{b2} J. Clerk Maxwell, A Treatise on Electricity and Magnetism, 3rd ed., vol. 2. Oxford: Clarendon, 1892, pp.68--73.
% \bibitem{b3} I. S. Jacobs and C. P. Bean, ``Fine particles, thin films and exchange anisotropy,'' in Magnetism, vol. III, G. T. Rado and H. Suhl, Eds. New York: Academic, 1963, pp. 271--350.
% \bibitem{b4} K. Elissa, ``Title of paper if known,'' unpublished.
% \bibitem{b5} R. Nicole, ``Title of paper with only first word capitalized,'' J. Name Stand. Abbrev., in press.
% \bibitem{b6} Y. Yorozu, M. Hirano, K. Oka, and Y. Tagawa, ``Electron spectroscopy studies on magneto-optical media and plastic substrate interface,'' IEEE Transl. J. Magn. Japan, vol. 2, pp. 740--741, August 1987 [Digests 9th Annual Conf. Magnetics Japan, p. 301, 1982].
% \bibitem{b7} M. Young, The Technical Writer's Handbook. Mill Valley, CA: University Science, 1989.
% \end{thebibliography}

% \vspace{12pt}
% \color{red}
% IEEE conference templates contain guidance text for composing and formatting conference papers. Please ensure that all template text is removed from your conference paper prior to submission to the conference. Failure to remove the template text from your paper may result in your paper not being published.

\end{document}